\documentstyle[psfig]{caosp}
\begin{document}
\htitle{Doppler imaging of $\kappa$~Psc}
\hauthor{N. Piskunov {\it et al.}}
\title{Multi-element Doppler imaging of $\kappa$~Psc}
\author{N.~Piskunov \inst{1}, H.C.~Stempels \inst{1}, 
        T.A.~Ryabchikova \inst{2}, V.~Malanushenko \inst{3},
        I.~Savanov \inst{3}}
\institute{Uppsala Astronomical Observatory, Uppsala, Sweden
\and       Institute of Astronomy, Moscow, Russia
\and       Crimean Astrophysical Observatory, Crimea, Ukraine}
\date{November 28, 1997}
\maketitle

\begin{abstract}
$\kappa$~Psc (HD 220825) is a typical Chromium Ap star that happens to have
optimal parameters for Doppler imaging (DI). Its short rotational period
of less then 2 days, rotational velocity of $\sim$40\,km/s, and a moderate
inclination of the rotational axis
put modest requirements on spectroscopic observations.
Anomalies of iron peak elements are clearly present, but
small enough to cause significant
deviations from model atmospheres with scaled solar abundances. We
applied DI to $\kappa$~Psc once before, determining the distribution of Cr
(Ryabchikova et al. \cite{rpdp}, hereafter Paper~I). However, due to strong
blending of Fe,
the image was based on two
short ($\sim$2\,\AA ) spectral intervals, dominated by Cr lines.
Since the first paper we obtained additional spectra and developed a new
code that allows to perform multi-element DI and thus to use larger spectral
interval(s).
We demonstrate the abilities of the
new code and present new maps of Cr and Fe. A much larger
time base allowed us to improve the rotational period of $\kappa$~Psc as well.

\end{abstract}

\section{Introduction}

Doppler imaging of stars (see e.g. Piskunov \& Rice 1993) has become a 
standard technique for studying the distribution of chemical elements on 
the surface of Ap stars. Up to now DI was limited to a single chemical 
element. In a few attempts to image several elements, maps were obtained 
sequentially, each time assuming that all elements but one are distributed 
homogeneously. Recent improvements in computer performance and
more efficient algorithms for solving radiative transfer allowed us to 
replace pre-computed tables of local line profiles with ``on the fly''
calculations of the emerging spectrum. The new code INVERS11, based on this 
approach, is not limited by the size of interpolation tables and can 
handle simultaneous imaging of multiple elements. In addition, we are able 
to use {\it blends} of different chemical elements for DI, which was not 
possible with the old codes.

We applied the new code to the SrCr star $\kappa$~Psc, classified A0p.
The selected spectral region around 5300\,\AA\ is dominated by lines of
neutral and ionized Cr and Fe lines, which give additional constraints for
the effective temperature. The rotational velocity is $\sim 38$~km\,s$^{-1}$
resulting in significant line blending.

\section{Observations and data reduction}

Our observational data include 25 high-resolution CCD spectra
($R~=~35000$, SNR~$\geq~200$) and consist of two sets. The first set was 
obtained with the Coud\'e spectrometer of the 2-m telescope of the Rozhen 
Astronomical Observatory (Bulgaria) in July and August 1993, and the 
second was observed with the Coud\'e spectrometer of the 2.6-m telescope 
of the Crimean Astrophysical Observatory in August and September 
1997. Our observations cover the spectral region 5285--5345 \AA, which is 
rich in Cr\,{\sc i} and {\sc ii} and Fe\,{\sc i} and {\sc ii} lines.

The Rozhen spectra were processed with the pcIPS software package 
(Smirnov \& Piskunov \cite{sp}), while the reduction of the CrAO spectra 
was done with the help of the SPE reduction package of S. Sergeev (CrAO).
Both packages include all standard procedures.
The continuum 
fitting was done with pcIPS for all spectra. Finally, small corrections 
to the the continuum level were applied using the synthetic spectrum of the 
star, which is particularly important for obtaining accurate line 
profiles in the spectrum of a star with substantial rotation.

\subsection{New period}

None of the rotational periods discussed in the literature, namely 1.420 
(Kreidl \& Schneider \cite{ks}), 1.412 (Kerschbaum \& Maitzen \cite{km}),
and 1.418 days (Ryabchikova et al. \cite{rpdp}) fitted our spectroscopic 
variations properly; spectra corresponding to close phases did not look 
similar. We therefore performed a new period search using the equivalent 
widths of Cr\,{\sc ii} lines in our spectra. This was done using the method by 
Cuypers (\cite{cu}), implemented by Pelt (\cite{pelt}). The best fit to all 
equivalent width variations was achieved with the period $P=1.409582 \pm 
0.000066$ days. The phases computed using this new period 
substantially improved the reproduction of the line shapes in close 
phases. Note that the change in the period does not 
influence the results of Paper~I because it was based on data
obtained in 14 consecutive nights.

\section{Line identification}

The initial line-list for the observed wavelength range
was extracted from the Vienna Atomic Line Database 
(Piskunov et al. \cite{pkr}). This line-list was then fine-tuned 
with the SME tool (Valenti \& Piskunov \cite{vp}) by comparing the 
synthetic spectrum to reference spectra of the Sun and Sirius (A1V), taken
from the NSO Solar Atlas (Kurucz et al. \cite{kz}), and for Sirius from the ING
archive at RGO.
Using SME, we first tuned the oscillator strengths of 
the most important lines of neutral atoms to match the NSO Solar Atlas.
Most of the ions have 
been tuned using observations for Sirius, although the spectral resolution 
and SNR are inferior to the NSO Solar Atlas data.
The resulting data consist of 25 spectra. We averaged the observations in 
close phases, further increasing the SNR, and performed DI 
using 11 rotational phase.

\begin{figure}[t]
\parbox[b]{65cm}{\psfig{figure=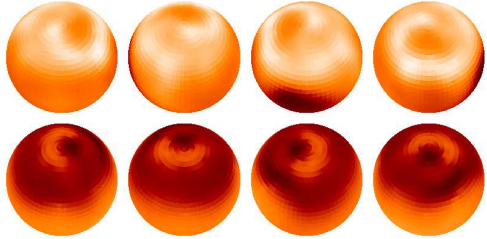,width=65mm}}
\hfill
\parbox[b]{53mm}{\caption{
Preliminary abundance maps of Cr (top) and Fe (bottom). The phases are,
from left to right 0.00, 0.25, 0.50 and 0.75. The abundance of Cr varies
between $-6.09$ and $0.27$, of Fe between $-6.15$ and $-1.81$.}
\label{fig:maps}}
\end{figure}

\subsection{Multi-element Doppler imaging}

We restricted the DI to the promising 5307--5320 \AA\ region. In other
regions our line-list seems to be incomplete and/or contain misidentified
lines.
In this region the new INVERS11 code was able to reproduce the spectra well
and to construct maps of Cr and Fe using a subset of 18 
spectral lines. The following parameters of $\kappa$~Psc resulted in the 
best fit: $v\sin i = 38$~km\,s$^{-1}$, $i = 70^\circ$,
$T_{\rm eff}= 9250$\,K,
$\log g = 4.5$, $v_{\rm micro}=2$~km\,s$^{-1}$. A preliminary abundance map is 
shown in Figure~\ref{fig:maps}.

We intend to publish finalised line-lists, the comparison between the
synthetic and observed spectra, and the data analysis in the near future
(Piskunov et al. \cite{pnew}).

\end{document}